# Correlations in the Binding Energy of Triexcitons and Biexcitons in Single CdSe/CdS Nanoplatelets Revealed by Heralded Spectroscopy

*Daniel Amgar,[1] Nadav Frenkel,[1] Dekel Nakar,[1] and Dan Oron[1,]\**

[1] Department of Molecular Chemistry and Materials Science, Weizmann Institute of Science, Rehovot 76100, Israel

KEYWORDS

Heralded spectroscopy, Auger recombination, nanoplatelets, SPAD arrays, multiexcitons, triexciton binding energy

ABSTRACT

Semiconductor nanoplatelets present reduced Auger recombination, giving rise to enhanced multiexciton emission. This virtue makes them good candidates to investigate higher-order carrier dynamics, allowing to extract important excitonic properties, such as biexciton and triexciton binding energies that highly influence applications involving high excitation fluxes. Here, we explore triexciton emission, emanating from single core/shell CdSe/CdS nanoplatelets. We apply heralded post-selection of photon triplets using an advanced home-built single-photon spectrometer in order to resolve the triexciton–biexciton–exciton–ground state cascaded relaxation both in time and spectrum, and unambiguously determine the triexciton relaxation route and interaction nature. The results show a characteristic blue shift of the biexciton and triexciton, pointing to repulsive multiexciton interaction in the nanoplatelets under study. The relatively small measured energy shift of the triexciton ($5.9 \pm 0.7$ meV) indicates that it recombines through the 1S bands rather than the 1P bands, in agreement with findings on other colloidal quantum dot systems. Most importantly, the strong correlation between the biexciton and triexciton binding energies, and the ability to tune them via control of the particle dimensions and composition, paves the way for developing emitters of nearly degenerate photon triplets.

INTRODUCTION

One of the key properties of semiconductor nanocrystals (SCNCs) is their ability to accommodate multiple excitons following an excitation event. Thus, multiexcitons are fundamental to SCNCs and affect many of their optoelectronic features, also through multi-carrier dynamics. The growing interest and motivation to explore multiexcitons are driven by applications demanding high excitation fluxes, such as lasing and displays, and a profound understanding of exciton–exciton interactions may improve the specifications for SCNCs when integrated into optoelectronic devices. More recently, there has been a growing interest in multiexciton emission cascades as potential sources of "on demand" quantum light (i.e., non-classical light such as entangled photon pairs), especially via the use of biexciton (BX) emission cascades. Indeed, while the BX state has been studied extensively in recent years[1,2], the triexciton (TX) state is yet to be fully understood in terms of recombination mechanisms, transition energy, and interaction with other excitons. As demonstrated in figure 1a, the TX state can relax through a cascaded emission process, emitting three consecutive photons (TX photon, BX photon, and single exciton photon, X), which are not only separated in time, but may also differ in energy due to many-body interactions. The energy difference between the TX and X photoluminescence (PL) peak energies is defined as the TX binding energy, similarly to the definition of the BX binding energy. Nearly all present investigations of TXs were performed through ensemble spectroscopy, including lifetime[3–6] measurements, spectrally-resolved[7] measurements and multidimensional spectroscopy[8,9]. Single-particle spectroscopy studies of the TX state were mostly performed at low temperature.[10,11] The few reports on room temperature TX spectroscopy dealt with the question of the nature of the emitting state. In a TX state, the two lowest-energy excitons occupy the S-like state and the third exciton occupies the P-like state.[3,7,12,13] This occupation in a strongly confined CdSe nanocrystal (NC) leads to two possible recombination routes for the TX cascaded emission, as illustrated in figure 1b; (i) Fast radiative recombination of the P exciton followed by two radiative recombination events of the S exciton (PSS), or (ii) Radiative recombination of an S exciton followed by fast thermalization of the charge carriers from the P band to the S band and then two subsequent radiative recombination events (SSS).[14] While ensemble spectroscopy of CdSe/ZnS quantum dots (QDs) showed that both routes contribute to recombination in the TX regime, it provided little information on the ratio between the two recombination mechanisms (branching ratio).[4] Recent work by Shulenberger et al. used spectrally-resolved third-order correlation measurements using color filters to determine the branching ratio in CdSe/CdS SCNCs and found that the SSS path is unequivocally the dominant one.[15] The fact that the emission lifetimes of all three photons emitted from a TX state are typically longer than the 100 ps time resolution of modern single-photon avalanche diode (SPAD) detectors enables the use of heralded spectroscopy[16], a technique in which one can post-select photon pairs, triplets, or higher states according to their time of arrival. Recent work on giant $CsPbBr_3$ QDs revealed similar trends to Shulenberger et al., using a single-photon sensitive spectrometer and heralded TX spectroscopy. Notably, these QDs operate in the weak confinement regime and exhibit very small (< 1 meV) BX and TX binding energies.[17]

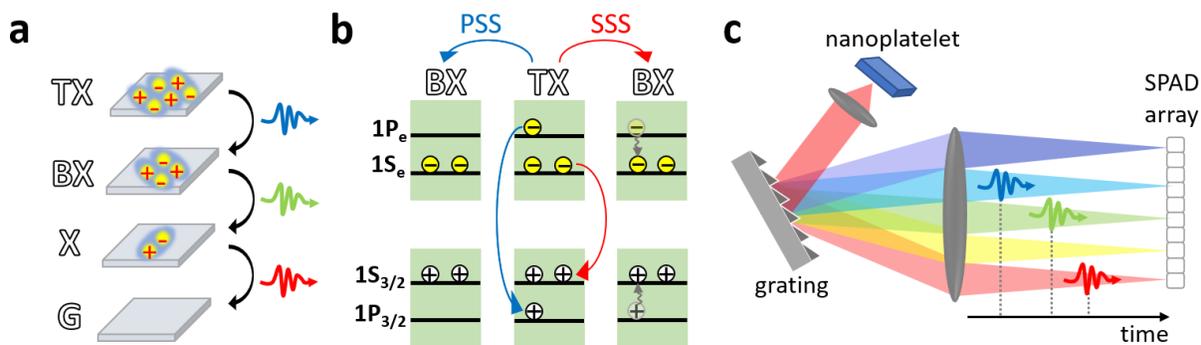

*Figure 1.* (a) Schematic of the triexciton (TX) emission cascade of a triply-excited nanoplatelet. The cascade features three consecutive transitions, defining the first emitted photon as a TX photon, the second emitted photon as a biexciton (BX) photon, and the third photon as a single exciton (X) photon. (b) Schematic of the possible TX recombination pathways. The blue arrow indicates the first step of the PSS route, where an electron and a hole from the 1P level recombine, leaving two S excitons. The red arrow indicates the first step of the SSS route, where an electron and a hole from the 1S level recombine, followed by fast thermalization, leaving again two S excitons (a BX state). In both pathways, the BX state then recombines through two S transitions. (c) Schematic of the spectroSPAD apparatus. The photoluminescence from a single nanoplatelet is entering a spectrometer. A blazed grating directs each single photon according to its wavelength to a different pixel of the linear SPAD detector, which is spectrally calibrated. Heralded post-selection is enabled through photon detection time-stamping and spectral resolution, resulting in the temporal and spectral isolation of TX events (demonstrated by three cascading photons arriving at different time delays within the same laser pulse).

One of the more intriguing potential applications of colloidal SCNCs is correlated or entangled few-photon sources.[18] While still very far from being fully explored, this would require precise control over exciton–exciton interactions. These interactions need to be sufficiently strong to induce correlations, implying at least a certain degree of quantum confinement, yet not too strong to cause overwhelming exciton–exciton annihilation. A particularly promising system in this respect is semiconductor nanoplatelets (NPLs), exhibiting strong confinement along one dimension and weak confinement along the other two. Indeed, NPLs have recently been shown to support multiple excitations due to suppressed Auger recombination, with the extent of coupling controlled by the NPL area and the nature of the interaction (repulsive or attractive) between excitons controlled by the composition.[19–21] However, research on the dynamics and kinetics of TX emission in those promising systems is still missing.

Here, we provide a direct and comprehensive characterization of the TX state emission in individual CdSe/CdS core/shell NPLs using heralded spectroscopy analysis adapted to characterize the TX binding energy and the relaxation route. One substantial advantage of the heralded spectroscopy technique over ensemble approaches is its ability to separate TX emission cascades from competing mechanisms such as emission from charged states. This enables, for example, elucidating the relationship between TX and BX binding energies, which we show to be highly correlated. Overall, in the ensemble of individually studied NPLs, we observe a blue-shift of both the TX and BX photons with respect to the X, indicating repulsive exciton–exciton interactions. We rationalize this repulsive nature by the inherent charge separation in the quasi type-II NPLs and the corresponding reduction in the overlap of the electron and hole wavefunctions. Finally, we point at the possibility of obtaining nearly degenerate photon triplets from such a cascade.

RESULTS AND DISCUSSION

The ~9 x 32 nm CdSe/CdS NPLs were synthesized as described in ref 20 (medium-sized). The CdSe cores (five monolayers) and CdSe/CdS core/shell (three CdS monolayers) NPLs were characterized via transmission electron microscope (TEM) and UV-vis absorption and PL measurements. The TEM images, spectra of the core and core/shell NPLs, and the corresponding technical details, are documented in section S1 and figure S1 in the supporting information (SI). Single core/shell CdSe/CdS NPLs were measured using a home-built setup, presented in figure 1c, termed spectroSPAD. The sample for measurement was prepared by drop-casting a diluted solution (x$10^4$ dilution) of the core/shell NPLs in hexane onto a glass coverslip. The repetition rate of the pulsed 470 nm excitation laser was 5 MHz and the power used was ~200 nW, corresponding to ~1.6 ± 0.2 photons emitted per NPL per pulse (see figure S2 in section S2 of the SI for more details about the saturation experiments). The spectroSPAD (figure 1c), was introduced in previous works of our group.[16,17,22] It combines a relatively high overall detection efficiency (~10% of the emitted photons are detected), low dark counts (~33 counts per second per detector pixel), and simultaneous spectrum and time detection capabilities at the single-photon level, by employing a high-performance linear SPAD array as a detector in a spectrometer configuration.[16] Briefly, an inverted microscope with a high numerical aperture objective is used to focus pulsed laser illumination on a single NPL, and to collect epi-detected PL. This signal is spectrally filtered from the excitation laser with a dichroic mirror and a dielectric filter, and imaged by a second lens. This image serves as the input for a spectrometer setup – a 4f system with a blazed diffraction grating at the Fourier plane. At the output image plane of the spectrometer, a monolithic linear pixelated SPAD array is placed, such that each pixel is aligned with the image of a different wavelength range. Single photon detections are time-tagged by an array of 64 time-to-digital converters. The apparatus is described in length in section S3 of the SI and in refs 16,17. The acquired time- and wavelength-tagged photon counts constitute an extremely informative dataset useful for higher-order cross-correlations of single photons. Using a dedicated MATLAB script, photon triplets, identified as events of a cascaded three-photon emission from a single NPL (figure 1a), are post-selected and analyzed through heralded spectroscopy, enabling to unambiguously determine the energy of each photon in the cascade and ascribe it to a certain transition in the radiative TX→BX→X→G (ground state) cascade.

Figure 2a displays the intensity ("blinking") trace as a function of time for an acquisition window of 200 seconds (total acquisition time was 2100 seconds), i.e., the sum of all detected photons in all pixels per 10 ms time-bins, of a representative single NPL. The blinking behavior is well-known and shows the alternating high and low emissivity states that are characteristic of SCNC emitters, which is also commensurate with a single-particle nature. Figure 2b shows a two-dimensional (2D) histogram of the photon counts per pixel per time delay from the preceding laser pulse, providing spectral–temporal information. The overlaid PL spectrum of the representative NPL, shown as a white dashed curve peaking at ~678 nm, was extracted by summing over all time delays (full horizontal binning). The lifetime curve of the NPL was extracted by summing over all energy values (full vertical binning; figure S3 in section S4 of the SI).

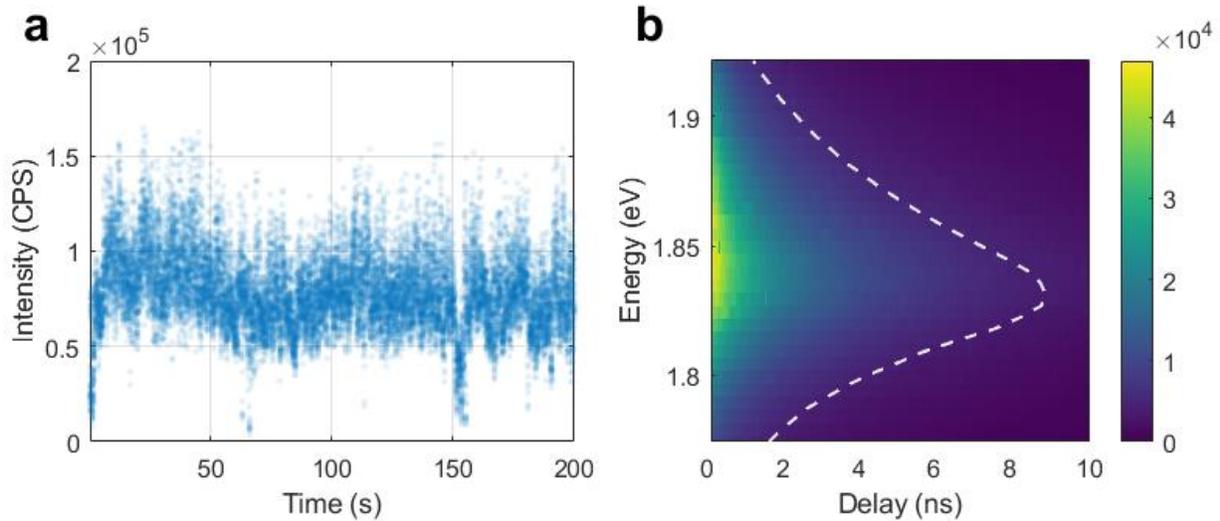

*Figure 2.* (a) Intensity trace for a 200-seconds acquisition time window for a representative NPL. (b) 2D energy–delay histogram of the detected photons. The color scale corresponds to the number of photons detected in each time delay per detector pixel, where each pixel corresponds to a narrow energy range. The white dashed curve is the PL spectrum of the representative NPL (generated by summing over all delay bins).

Figure 3a shows the normalized second-order correlation ($g^{(2)}$) versus the delay between pair detections. $g^{(2)}(0)$ is the probability to detect photon pairs almost simultaneously (i.e., within the same excitation pulse). The reduction of the magnitude of the peak around zero time delay reflects the lower probability of detecting these photon pairs and quantifies the photon antibunching. Unlike strongly confined QDs, the probability of emitting multiple photons per excitation pulse in single CdSe/CdS NPLs (and similar systems) is far from zero, as has already been previously demonstrated and discussed.[19–21,23,24] The NPL's weakly quantum-confined, larger lateral dimensions with respect to the strongly confined thickness allows for a higher lateral separation of excitons, substantially suppressing the nonradiative Auger recombination process. Thus, the $g^{(2)}(0)$ value of the NPL in figure 3a of $0.765 \pm 0.001$ is reasonable and agrees well with previous observations.[20] In addition, figure 3b shows a 2D representation of the third-order photon correlation ($G^{(3)}$), i.e., the probability to detect photon triplets as a function of the time delay between detections (binned to 200 ns, the time between consecutive pulses). The $(\tau_1, 0)$, $(0, \tau_2)$, and $(\tau_1 = \tau_2)$ pixelated lines represent two photons detected simultaneously (photon pairs) and a third photon detected within another laser pulse. The other areas at $(\tau_1, \tau_2)$, where $\tau_1, \tau_2 \neq 0$ and $\tau_1 \neq \tau_2$, represent three uncorrelated photon detection events occurring at different excitation pulses, used here for normalization. $G^{(3)}(0,0)$ is extracted by the photon counts in the central datapoint at $(0,0)$, which represents the probability to detect three photons within the same excitation pulse. For the representative NPL shown in figure 3b, the estimated normalized $g^{(3)}(0,0)$ is $0.571 \pm 0.009$. It is calculated by dividing the photon counts in the central datapoint by the average photon counts in all $(\tau_1, \tau_2)$ datapoints (where all three detections are time-distant by at least one pulse). To alleviate optical crosstalk artifacts, the $G^{(3)}$ data were calculated by including only photon triplets that were five or more detector pixels apart and arrived after a time threshold of 0.2 ns (more details on the applied crosstalk reduction is provided in ref 17). A plot of $g^{(3)}(0,0)$ versus $[g^{(2)}(0)]^2$ of all measured NPLs, shown in figure 3c, assures that higher-order moments indeed align with lower orders, in agreement with the binary collision model and in general agreement with the results obtained without spectrally resolving the emission, using a fiber beamsplitter.[20] Notably, the mean value of the antibunching [$g^{(2)}(0)$] for all measured NPLs, $0.866 \pm 0.007$, is higher than in previous reports.[20,24] This increase can be associated with saturation effects in accordance with the work by Nair et al.[25] They report on the excitation power dependence of $g^{(2)}(0)$ in CdSe-like NCs, which increases linearly for small $g^{(2)}(0)$ values and then saturates at high excitation powers due to reduced quantum yield (QY) of higher multiexcitons.[25] As mentioned above, our measurements were performed slightly above saturation, which may explain the higher $g^{(2)}(0)$ values. Furthermore, a previous work on these NPLs[20] showed a small but statistically significant downward deviation from the model, which points at an accelerated Auger recombination of photon triplets. Interestingly, an opposite deviation is observed is this case. This could be an artifact of post-selection of NPLs with a relatively large number of TX counts. Presumably, this selectivity is toward larger NPLs with thicker CdS shells, since they have a higher TX emission rate. A thicker shell promotes type-II band alignment since the electron is effectively more localized in the CdS shell. The charge separation, which is probably more prominent is this case, may increase the repulsion among multiexcitons, possibly somewhat slowing down nonradiative TX recombination relative to the case where an attractive three-body interaction is observed.[20] Further details about the $g^{(3)}(0,0)$ versus $[g^{(2)}(0)]^2$ correlation and binary collision model are provided in ref 20.

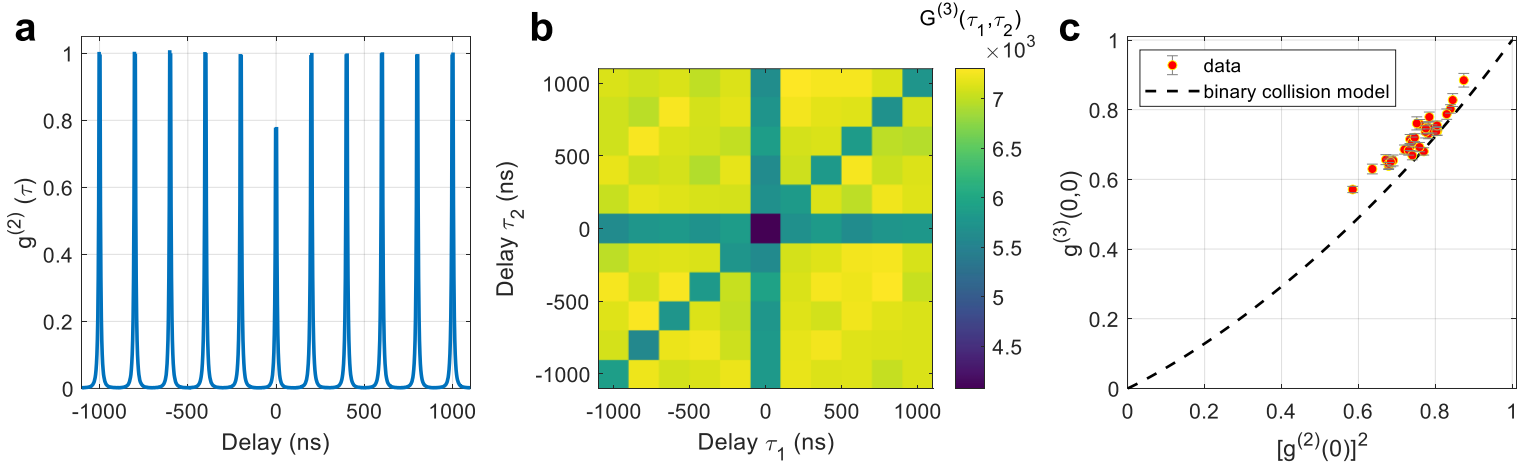

*Figure 3.* (a) Second-order temporal photon correlation [$g^{(2)}(\tau)$] of the same NPL analyzed in figure 2 as a function of the time delay (2.5 ns bins). (b) Third-order temporal photon correlation [$G^{(3)}(\tau_1,\tau_2)$, 200 ns bins]. (c) Third-order correlation at time delays of zero between the three detections [$g^{(3)}(0,0)$] versus second-order correlation at zero time delay squared $[g^{(2)}(0)]^2$ for all measured NPLs. The black dashed line represents $g^{(3)}(0,0)$ calculated through the binary collision model (see details in ref 20). A magnified view of the datapoints appears in figure S4 in section S5 of the SI.

Using heralded spectroscopy, we resolve the different steps of the emission cascade both energetically and temporally. We first consider the exciton–exciton interaction energies in the BX and TX states. The BX binding energy [$\varepsilon_b(\text{BX})$] is defined here as the difference between the BX and the X emission energies, determined by the PL peak centers: $\varepsilon_b(\text{BX}) = E_{BX} - E_X$. Thus, a positive $\varepsilon_b(\text{BX})$ indicates repulsive interaction and a negative $\varepsilon_b(\text{BX})$ indicates binding. Previous work from our group by Lubin et al.[16], and other works [22,26], demonstrated spectral correlations between the BX and the X photons in CdSe/CdS/ZnS QDs through BX heralded spectroscopy. In ref 16, the $\varepsilon_b(\text{BX})$ was determined with sub-meV precision indicating an attractive exciton-exciton interaction that is positively correlated with the level of quantum confinement of the QDs. The potential of heralded spectroscopy to measure higher-order photon correlations due to the large number of pixels in the SPAD detectors inspired us to leverage it toward measuring photon triplets in a NC system which supports this multiexciton level in terms of stability and QY, such as quasi-type II core/shell NPLs. This is particularly interesting with respect to the possibility of emitting multiple degenerate photons in a cascade, potentially enabling formation of states such as heralded entangled pairs by schemes such as time reordering.[27] Figures 4a and 4b present TX heralded spectroscopy analysis from a single-NPL measurement. Typically, the BX emission rate is larger than the TX emission rate by two orders of magnitude, emphasizing the challenge in capturing a sufficiently large number of TX events to obtain adequate statistics. In this NPL case, ~823,000 photon pairs and ~2,300 photon triplets have been recorded. Figure 4a shows the TX (blue), BX (green), and X (red) PL spectra, along with the normalized PL spectrum of all photon detections (gray area), without any post-selection. We note that although the energy axis is centered on the spectral range of the NPL emission, the spectroSPAD detector spans a broader spectral window of approximately 100 nm (~300 meV), as shown in figure S5 in section S6 of the SI. The spectra are fitted by a Cauchy-Lorentz distribution (dashed lines in figure 4a), peaking at $1.839 \pm 0.001$ eV, $1.834 \pm 0.001$ eV, and $1.832 \pm 0.001$ eV, for the TX, BX and X, respectively, and present a blue shift of the BX and a further blue shift of the TX emission. This result exemplifies the power of the heralded spectroscopy technique, which analyzes post-selected photon triplets, and directly extracts information on the first three excited states. Thereafter, similarly to the BX case, the TX binding energy $[\varepsilon_b(TX)]$ can be estimated from the difference between the TX and X peak energies as follows: $\varepsilon_b(TX) = E_{TX} - E_X$. For the representative NPL, $\varepsilon_b(BX) = 2.3 \pm 1.2$ meV and $\varepsilon_b(TX) = 7.7 \pm 1.1$ meV. Previous studies have assigned a BX blue shift to a stronger Coulomb repulsion that occurs in type-II or quasi-type-II (as is the case here) QDs, where electrons and holes are spatially separated, which is in good agreement with the NPLs under study.[28]

Next, we use the sub-ns time resolution of heralded spectroscopy to study the kinetics of the different relaxation steps of the TX state. Figure 4b shows the lifetime curves of the TX, BX, and X states and the corresponding monoexponential fit (dashed lines). As expected, the TX has the shortest monoexponential lifetime, of ~$0.62 \pm 0.02$ ns, followed by the BX lifetime of ~$1.3 \pm$ 0.1 ns, and then the X lifetime of ~4.1 ± 0.4 ns. This ascending order of time constants agrees with the more rapid recombination expected from higher excited states due to both the higher degeneracy of emitting states and the enhanced rate of Auger recombination at larger carrier numbers within the NCs.[5,29] To show the generality of these results, other examples of single NPLs analysis are presented in figure S6 in section S7 of the SI. To validate that all measurements originate from single NPLs, rather than clusters, one can seek for nearly monoexponential dynamics of the BX and TX lifetime curves, as a pair or a larger cluster of NPLs would have shown bi- or triexponential dynamics, as reported in previous works.[17,30] In the case of the representative NPL shown in figure 4b, $R^2 = 0.997$ and $R^2 = 0.972$ for the monoexponential fits of the TX and BX lifetime decays, respectively, in strong agreement with the assumption of single exponential decay. To further verify the single-particle nature of the measured NPLs, a time gating assay of the $g^{(2)}$ function has been applied to each measurement. Photons arriving at times outside a certain time gate (less than 6 ns and more than 35 ns in this case) are filtered out, practically containing all of the multiexciton emission, and only late-arriving photons, expected to originate from singly excited NCs (therefore single-photon emitters), are used to construct the *gated $g^{(2)}(\tau)$* curve. NPL measurements who met the criterion $g^{(2)}(0) < 0.5$ were considered single, according to the formula $g^{(2)}(0) = 1 - \frac{1}{n}$ for an ensemble of n single-photon emitters, and taken for further analysis.[31] These amounted to 31 single NPLs; two NPLs passed $g^{(2)}(0) \leq 0.3$, seventeen additional NPLs passed $g^{(2)}(0) \leq 0.4$, and twelve additional NPLs passed $g^{(2)}(0) < 0.5$. For instance, the *gated $g^{(2)}$* of the NPL analyzed in figure 3 is 0.26 ± 0.09 (see figure S7 in section S8 of the SI). Figure 4c presents the distribution of lifetime values among all measured individual NPLs. The mean lifetime values are 0.51 ± 0.02 ns, 1.03 ± 0.04 ns, and 3.09 ± 0.15 ns for the TX (blue), BX (green), and X (red) states, respectively. In the heralded analysis process, the photons assigned to each state are time-gated according to their delay, in order to address dark counts (DC) and inter-pixel optical crosstalk artifacts that may skew the results, as follows; TX photons are gated between –0.5 and 8 ns from the laser pulse, BX photons are gated between 0.3 and 8 ns from the first detection (TX), and X photons are gated between 0.3 and 35 ns from the second detection (BX), depicted in figure 4c by vertical dashed, dashed-dotted, and dotted lines, respectively. The gating of TX photons aims to reduce the number of DC falsely counted as part of photon triplets, and the gating of the BX and X photons aims to reduce the number of crosstalk photons and DC that falsely counted as part of photon pairs and triplets. The lower limit of the BX and X time-gate filters out most of the crosstalk photons triggered by the preceding detection, while the upper limit filters DC, as previously shown in ref 32.

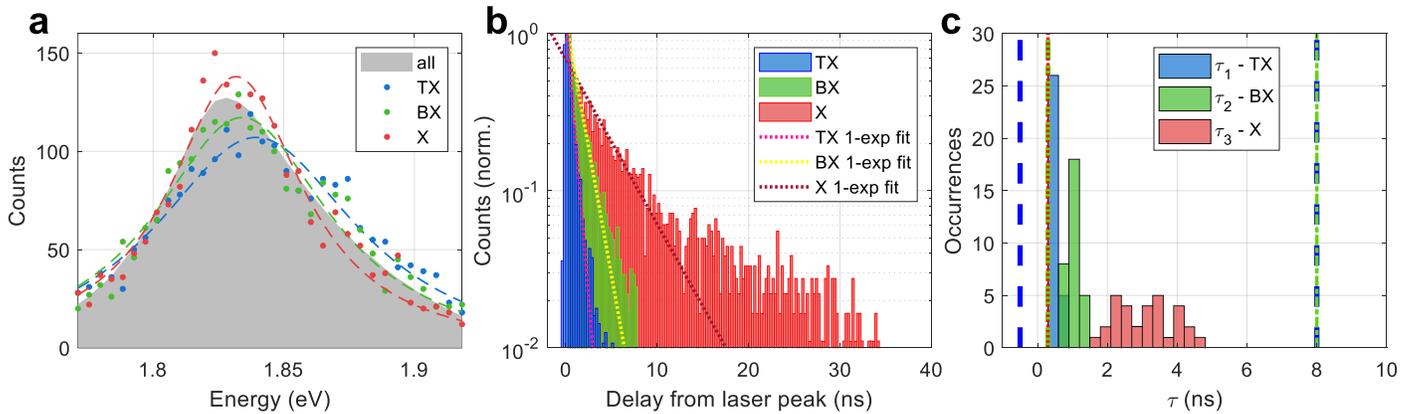

*Figure 4.* (a) Single NPL spectra of the TX (blue), BX (green), and X (red) photons. The blue, green, and red dashed lines are fits of the TX, BX, and X spectra, respectively, to a Cauchy-Lorentz distribution. The gray area corresponds to all detected photons, without heralded post-selection. (b) Lifetime histograms of the single NPL along with monoexponential fits. The delay time of the TX is from the laser peak, and the delay times of the BX and X are from the preceding arrival times of the TX and BX, respectively. (c) Distribution of the time constants ($\tau$) extracted from the monoexponential fits for the TX, BX, and X photons for all measured NPLs. The blue dashed vertical lines set the time-gate for the TX (-0.5 to 8 ns). The green dashed-dotted lines set the time-gate for the BX (0.3 to 8 ns) and the dotted red line sets the first time-gate of 0.3 ns for the X (the upper time gate, 35 ns, is not indicated in this plot).

As mentioned above, $\varepsilon_b(BX)$ can be calculated from the difference between the BX and X spectral peaks ($\varepsilon_b(BX) = E_{BX} - E_X$). Similarly, $\varepsilon_{TX}$ can be estimated from the difference between the TX and X spectral peaks ($\varepsilon_b(TX) = E_{TX} - E_X$). Our direct measurement of cascaded TX emission allows extracting both $\varepsilon_b(TX)$ and $\varepsilon_b(BX)$. Figure 5 depicts the ensemble results of the $\varepsilon_b(TX)$ and $\varepsilon_b(BX)$ measured from 31 single NPLs. Such measurement and the following heralded analysis allow to gain statistically significant insights regarding the energetics of exciton-exciton interactions in a higher multiexciton level. Figure 5a shows $\varepsilon_b(TX)$ and $\varepsilon_b(BX)$ as a function of the spectral position of the exciton peak, estimated through heralded analysis, along with a linear fit of each (black and red lines respectively). We find that $\varepsilon_b(TX)$ is larger than $\varepsilon_b(BX)$, that is, the mean TX energy is blue-shifted relative to the BX, which in turn is blue-shifted relative to the X. The bluer TX and BX with respect to the X, which translates to positive binding energies (according to the $\varepsilon_b(BX)$ and $\varepsilon_b(TX)$ convention used here), is related to a repulsive interaction between the excitons. Notably, the slope of the TX repulsion curve as a function of the X peak energy is more than three times that of the BX curve, and the two cross each other very close to the horizontal axis (zero binding energy). This implies that the spectral shift of the TX emission from the BX is nearly equal to that of the BX from the X, which supports the SSS relaxation mechanism of the TX (where both TX and BX photons decay from the S state). In a PSS mechanism, the TX photon would have shown a much larger blue shift due to the 1P-1S energy splitting (expected to be above 100 meV for this system[14,15,33]), which in principle could have been detected by our spectroSPAD system. Figure 5b further presents histograms of $\varepsilon_b(TX)$ and $\varepsilon_b(BX)$ for all measured NPLs, with mean values of $5.9 \pm 0.7$ meV and $1.6 \pm 0.4$ meV, respectively. Notably, there is a negative correlation between the spectral position of the X peak and the TX and BX binding energies. Lower energy excitons are usually associated with thicker shell NPLs, in which the electrons are more confined to the CdS shell, whereas the holes are in the core in all cases (type-II-like). Thus, a plausible result is a reduced Coulomb attraction between the excitons and thus more positive TX and BX binding energies (weaker binding), as shown in previous reports.[16,26,34] Similarly, our results probably originate from the shell thickness variation among the measured NPLs, which points at decreased exciton-exciton repulsion for thinner-shell NPLs. The strong correlation between the TX and BX binding energies is further demonstrated in figure S8 in section S9 of the SI, revealing few NPLs for which the binding energies are close to zero. One interesting consequence of this correlation is that it is likely possible, using a slightly thinner CdS shell (promoting type-I band alignment), to obtain NPLs exhibiting near zero TX and BX binding energies even in the strongly confined regime, implying the emission of multiple nearly degenerate photons in a cascaded emission process.

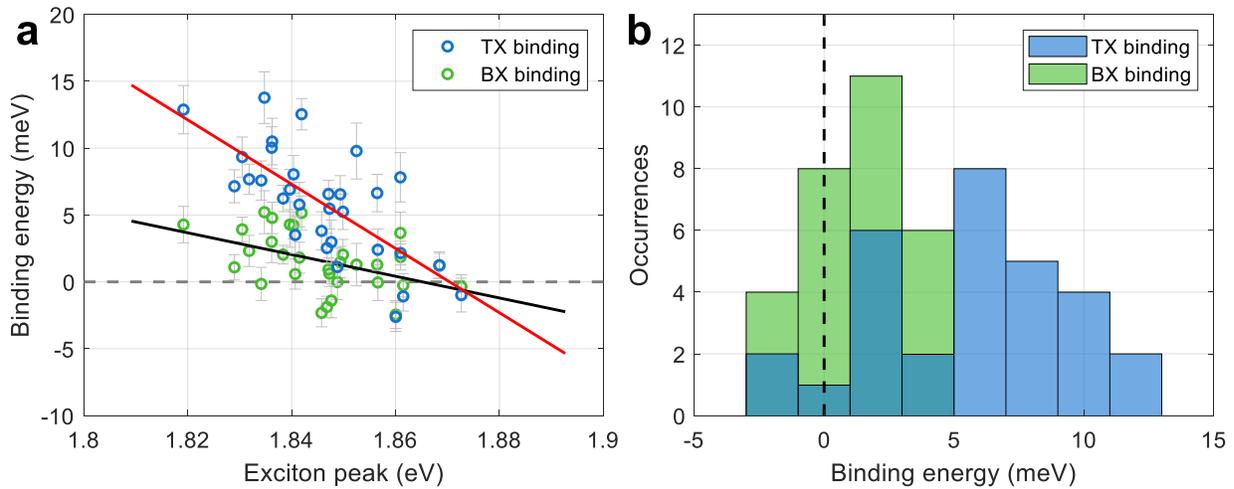

*Figure 5.* (a) TX (blue) and BX (green) binding energies calculated from the heralded analysis correlated with the X spectral peak. Red and black solid lines are linear fits of the TX and BX binding energies, respectively. The horizontal gray dashed line corresponds to a zero spectral shift (binding energy of zero) of the TX and BX. (b) Histograms of the TX (blue) and BX (green) binding energies with 2 meV bins. The vertical black dashed line marks the point of zero binding energy, i.e., no interaction between the excitons. The mean TX and BX binding energies are $5.9 \pm 0.7$ meV and $1.6 \pm 0.4$ meV, respectively.

CONCLUSIONS

We present a direct single-particle measurement of the TX binding energy in core/shell CdSe/CdS NPLs. A home-built single-photon spectrometer, integrating novel SPAD array technology, allows us to apply heralded spectroscopy and post-select photon triplets, which are then directly characterized by extracting the TX, BX, and X spectra. We have measured several tens of NPLs and the ensemble results revealed a trend of slightly blue-shifted BX and an even more blue-shifted TX, translating into repulsive TX binding energies. We have found that as the X energy increases, the TX is less repulsive, rationalized by variability in the CdS shell thickness of different single NPLs. Accordingly, NPLs with thinner shells are characterized more by type-I band alignment (confinement of the charge carriers in the CdSe core), which results in higher binding (less positive TX binding energy). Moreover, the small blue shift of the observed TX photons supports the conclusion of Shulenberger et al., that SSS band-edge recombination pathway is the dominant one.[15]

This direct technique, based on heralded analysis, constitutes a platform to multiexciton characterization in the single-particle level in many SCNC systems and allows for spectral selection of the BX and TX photons. This spectral control opens the possibility of utilizing these NPLs as a source of "on demand" pairs or triplets of spectrally indistinguishable photons.

## ASSOCIATED CONTENT

**Supporting Information**. Characterization of the CdSe/CdS NPLs by transmission electron microscope and absorption and photoluminescence spectra; details of the saturation estimation assay; details of the SPAD array-based single-photon spectrometer (spectroSPAD) apparatus; single-particle lifetime analysis; magnified plot of higher-order photon correlation; heralded spectra with full spectral range; additional examples of single-particle measurements; time gating of the second-order photon correlation function for a representative NPL; correlation of triexciton and biexciton binding energies.

## AUTHOR INFORMATION

**Corresponding Author**

*Dan.Oron@weizmann.ac.il

**Author Contributions**

D.A. synthesized the nanocrystals, performed the experiments and the data analysis, and wrote the manuscript. N.F. and D.N. assisted in data analysis. D.O. supervised the work. The manuscript was written through contributions of all authors. All authors have given approval to the final version of the manuscript.

**Funding Sources**

This research was supported by the Israel Science Foundation (ISF), and the Directorate for Defense Research and Development (DDR&D), grant No. 3415/21. D.A. gratefully acknowledges support by the VATAT Fellowship for female PhD students in Physics/Math and Computer Science. D.O. is the incumbent of the Harry Weinrebe professorial chair of laser physics.

**Notes**

The authors declare no competing financial interest.

ACKNOWLEDGMENT

The authors thank Dr. Gur Lubin and Dr. Ron Tenne for developing the spectroSPAD system and analysis script and establishing its use, Edoardo Charbon, Claudio Bruschini, and Michel Antolovic for the help with development of the detection hardware, and Gaoling Yang for helping with the synthesis of the NPLs cores of the used sample.

# Supporting Information: Correlations in the Binding Energy of Triexcitons and Biexcitons in Single CdSe/CdS Nanoplatelets Revealed by Heralded Spectroscopy

*Daniel Amgar,[1] Nadav Frenkel,[1] Dekel Nakar,[1] and Dan Oron[1,]\**

[1] Department of Molecular Chemistry and Materials Science, Weizmann Institute of Science, Rehovot 76100, Israel

# S1: Characterization of the CdSe/CdS nanoplatelets

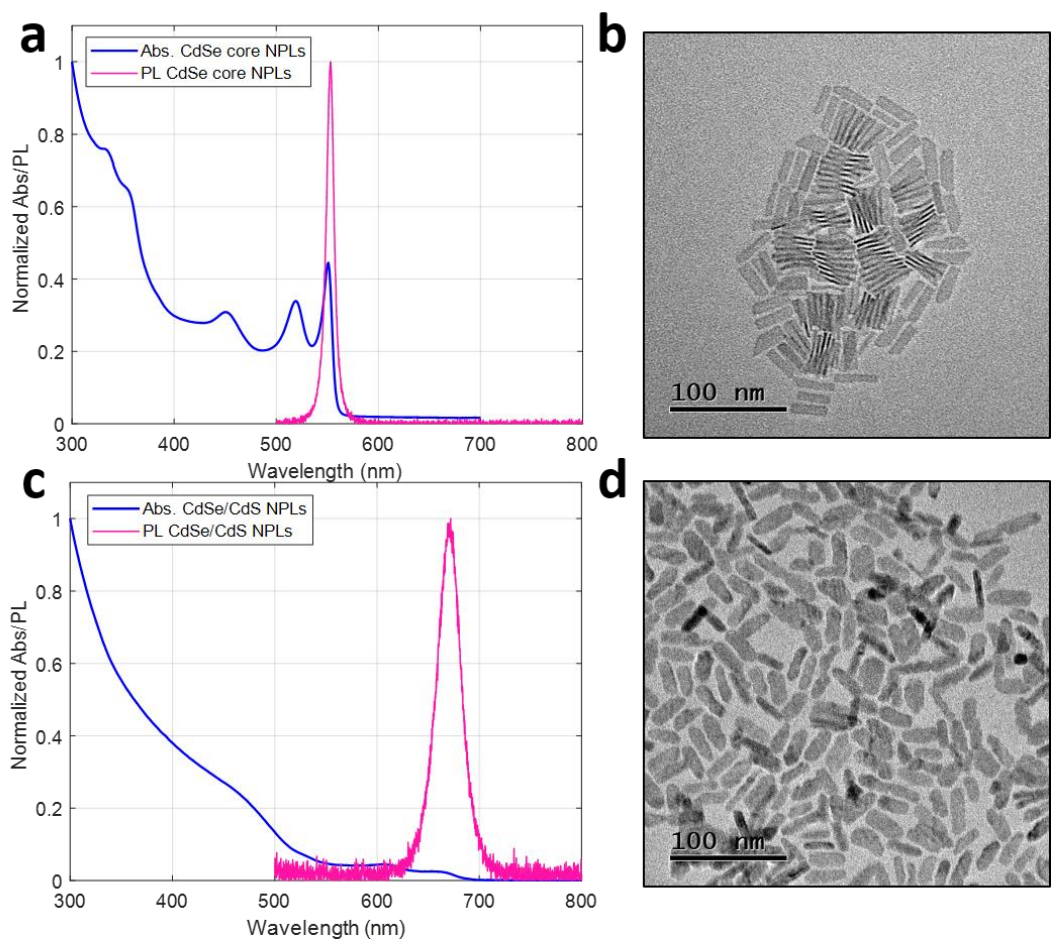

*Figure S6.* (a) Normalized absorption (Abs, blue) and photoluminescence (PL, pink) of the 5-monolayer CdSe core NPLs. (b) A transmission electron microscope (TEM) image of the CdSe core NPLs. (c) Normalized absorption (blue) and PL (pink) of the CdSe (5 monolayers)/CdS (3 monolayers) core/shell NPLs. (d) A TEM image of the CdSe/CdS core/shell NPLs.

*Characterization methods.* TEM images were taken on a JEOL 2100 TEM equipped with a LaB6 filament at an acceleration voltage of 200 kV on a Gatan US1000 CCD camera. UV-vis absorption spectra were measured using a UV-vis-NIR spectrometer (V-670, JASCO). Photoluminescence spectrum was measured using USB4000 Ocean Optics spectrometer excited by a fiber-coupled 407 nm LED in an orthogonal collection setup.

# S2: Saturation assay

In order to determine the suitable laser power to be used in the experiments, a single-particle saturation measurement was performed, as described at length in the supporting information (SI) of ref 1. Briefly, a single particle was illuminated with increasing laser power from ~35 nW to ~500 nW, in 15 equally-spaced power steps of ~30 nW, where each power step was measured for 10 seconds. Then, the power was decreased from the maximal power back to the starting power in a similar stepwise manner so that overall the particle was measured at each power for 20 seconds (10 seconds in the ascent and another 10 seconds in the descent). A saturation curve was constructed, as shown in figure S1 and described in the SI of refs 1,3. The curve was fitted to a saturation function, extracting a nominal saturation power of 132 nW, and the actual power used in all measurements was 200 nW (nominally 205 nW, due to a 5 nW deviation between the calibrated power in the measurement of figure S1 to the actual power measured at the back entrance of the microscope with a power meter), which is beyond saturation. This power was used in order to enable measuring a sufficient number of photon triplets, assuming a negligible number of photon quadruplets. The probability to absorb at least $n$ photons per excitation pulse can be estimated from the Poissonian distribution, giving probabilities of ~79%, ~46%, ~20%, and ~7% for absorbing at least one, two, three, and four photons, respectively. The average number of absorbed photons per pulse ($\langle N \rangle$) can be calculated by the power used ($I_{used}$) divided by the extracted saturation power ($I_{sat}$): $I_{used}/I_{sat}$, which in this case is ~$1.6 \pm 0.2$. We note that due to the non-negligible $g^{(2)}(0)$, this saturation is that of emitted photons, rather than absorbed photons, meaning that at $I_{sat}$, more than one photon is absorbed per pulse.

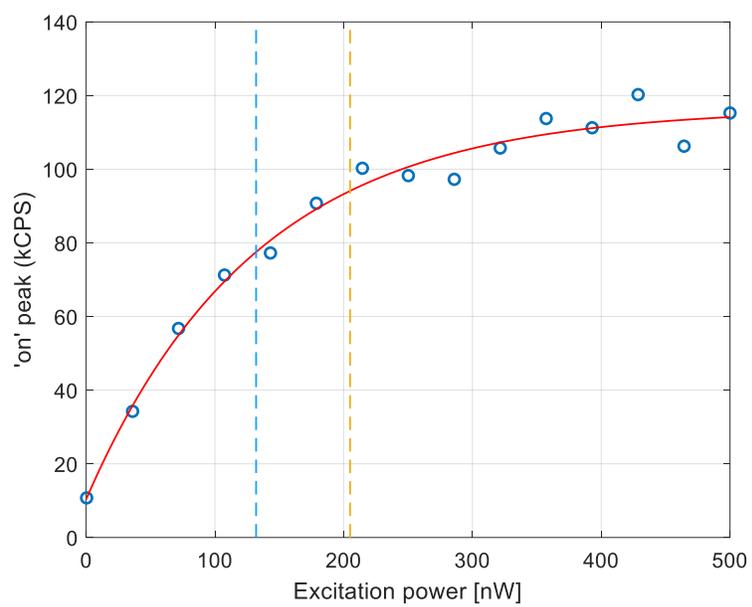

*Figure S7.* Saturation curve measured for a single nanoplatelet. The orange dashed line in 205 nW indicates the nominal power used in this measurement and the blue dashed line at 132 nW indicates the saturation power extracted from the fit to a saturation function (red curve).

## S3: Apparatus

The spectroSPAD system is practically a home-built single-photon spectrometer. A 470 nm pulsed laser (70 ps, LDH-P-C-470B, PicoQuant) with repetition rate of 5 MHz enters the back of an inverted optical microscope (Eclipse Ti-U, Nikon), reflected by a 484 long-pass dichroic mirror (FF484-FDi02-t3, Semrock), and focused on a single particle in the sample through a high NA x100 oil-immersion objective lens (1.3 NA, Nikon). The epi-detected photoluminescence (PL) emitted from the single particle is then collected via the same objective, passes the dichroic mirror and a 473 nm long-pass emission filter (BLP01-473R, Semrock) in order to filter out the 470 nm excitation laser. The image plane of the microscope is the input of a Czerny-Turner spectrometer based of a 4f system (AC254-300-A-ML and AC254-100-A-ML, Thorlabs) and a blazed diffraction grating (53-*-426R, Richardson Gratings). The output of the spectrometer is coupled to a monolithic linear single-photon avalanche diode (SPAD) array detector (synchronized with the laser), which is composed of 512 pixels with a pixel pitch of 26.2 μm (spectral resolution is ~1.7 nm), out of which 64 pixels are connected to a time-to-digital converters (TDCs) implemented by a field-programmable gate array (FPGA). This module registers the time stamp and pixel of each detected photon so that finally a list of photons along with their pixel addresses and arrival times (given by the pulse number after which the photon has arrived and the time delay from the beginning of that pulse). Due to excessive dark counts in pixel 34, we excluded it from the analysis. Further details are found in refs 1,2.

# S4: Lifetime analysis of a single nanoplatelet (NPL)

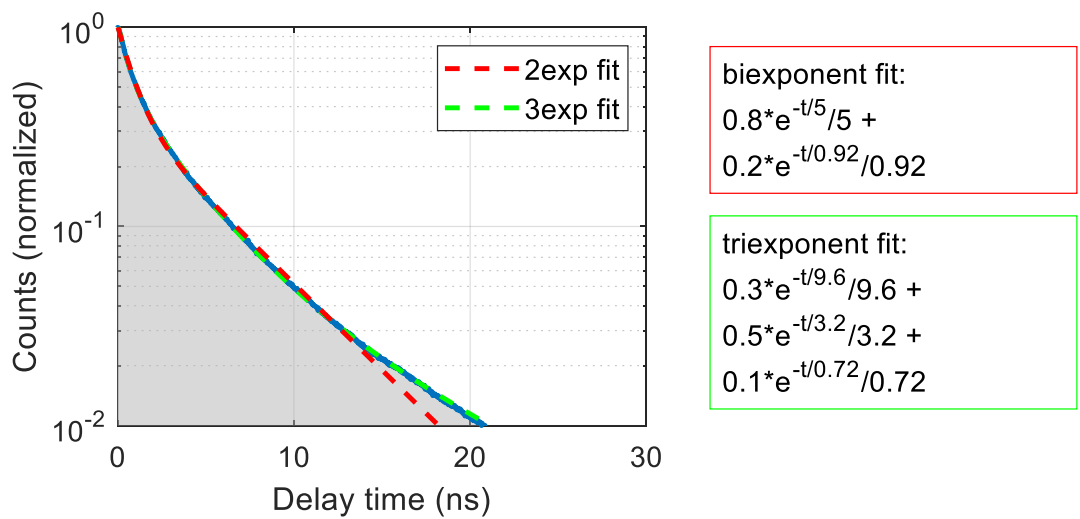

*Figure S8.* Lifetime curve (blue solid line) of a single NPL extracted by full vertical binning of figure 2b of the main text, along with a biexponential (dashed red line) and triexponential (dashed green line) fits. The respective fitting results are presented on the right side, yielding lifetimes of 5 ns and 0.92 ns from the biexponential fit and lifetimes of 9.6 ns, 3.2 ns, and 0.72 ns from the triexponential fit.

# S5: Third- versus second-order photon correlation – magnified image

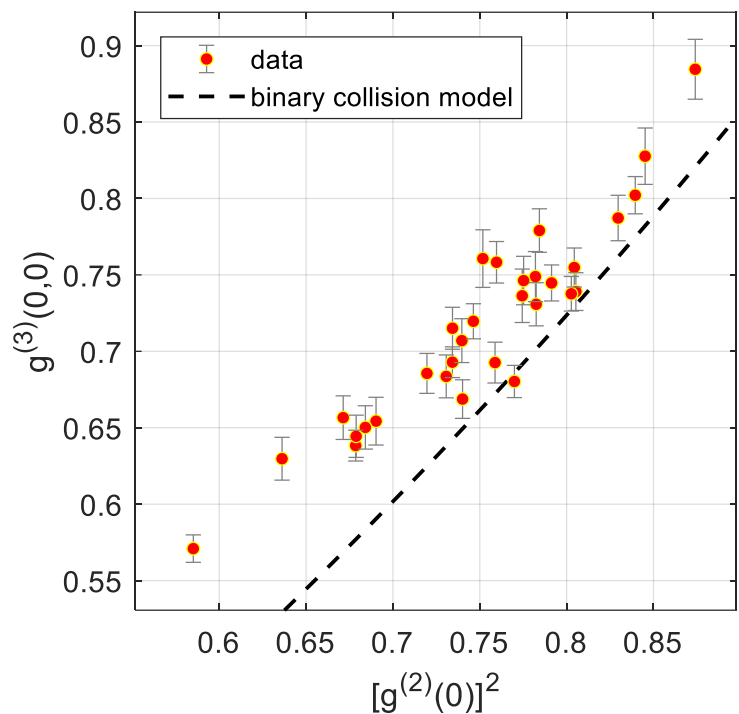

*Figure S9.* A magnified view of the datapoints in figure 3 of the main text. Third-order correlation at time delays of zero between the three detections [$g^{(3)}(0,0)$] versus second-order correlation at zero time delay squared $\{[g^{(2)}(0)]^2\}$ for all measured NPLs. The black dashed line represents $g^{(3)}(0,0)$ calculated through the binary collision model (see details in ref 4).

# S6: Heralded spectra spanning all of the detector spectral range

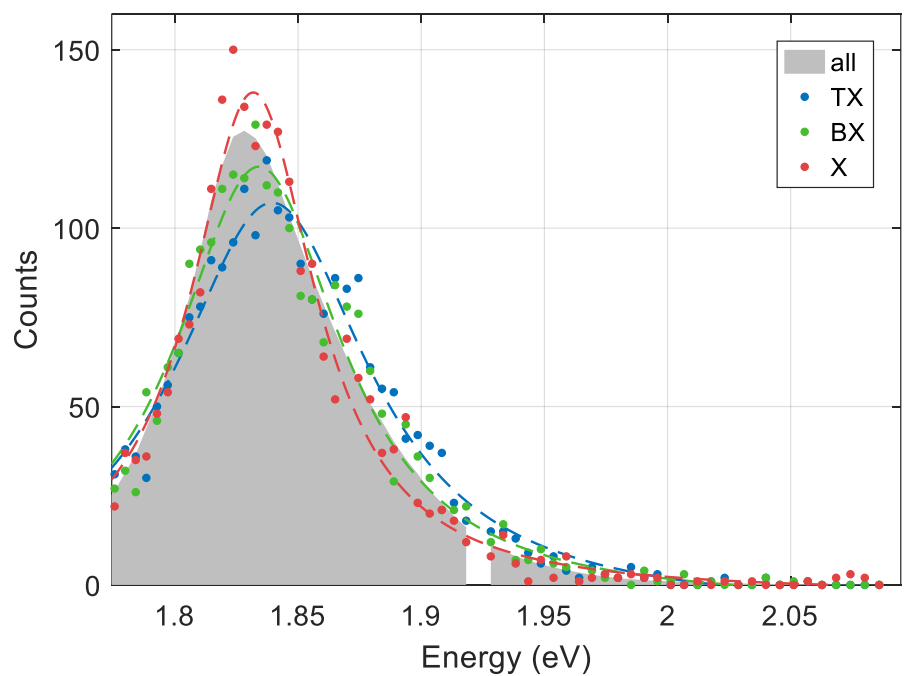

*Figure S10.* Single NPL spectra of the triexciton, TX (blue), biexciton, BX (green), and exciton, X (red) photons. The blue, green, and red dashed lines are fits of the TX, BX, and X spectra, respectively, to a Cauchy-Lorentz distribution. The gray area corresponds to all detected photons, without heralded post-selection, normalized to match the TX counts. The nullified data at ~1.93 eV is due to the removal of the noisy pixel. This plot corresponds to figure 4a of the main text, displaying a zoomed-out view that reveals the full spectral range detectable by the spectroSPAD system.

# S7: Additional examples of single-nanoplatelet analysis

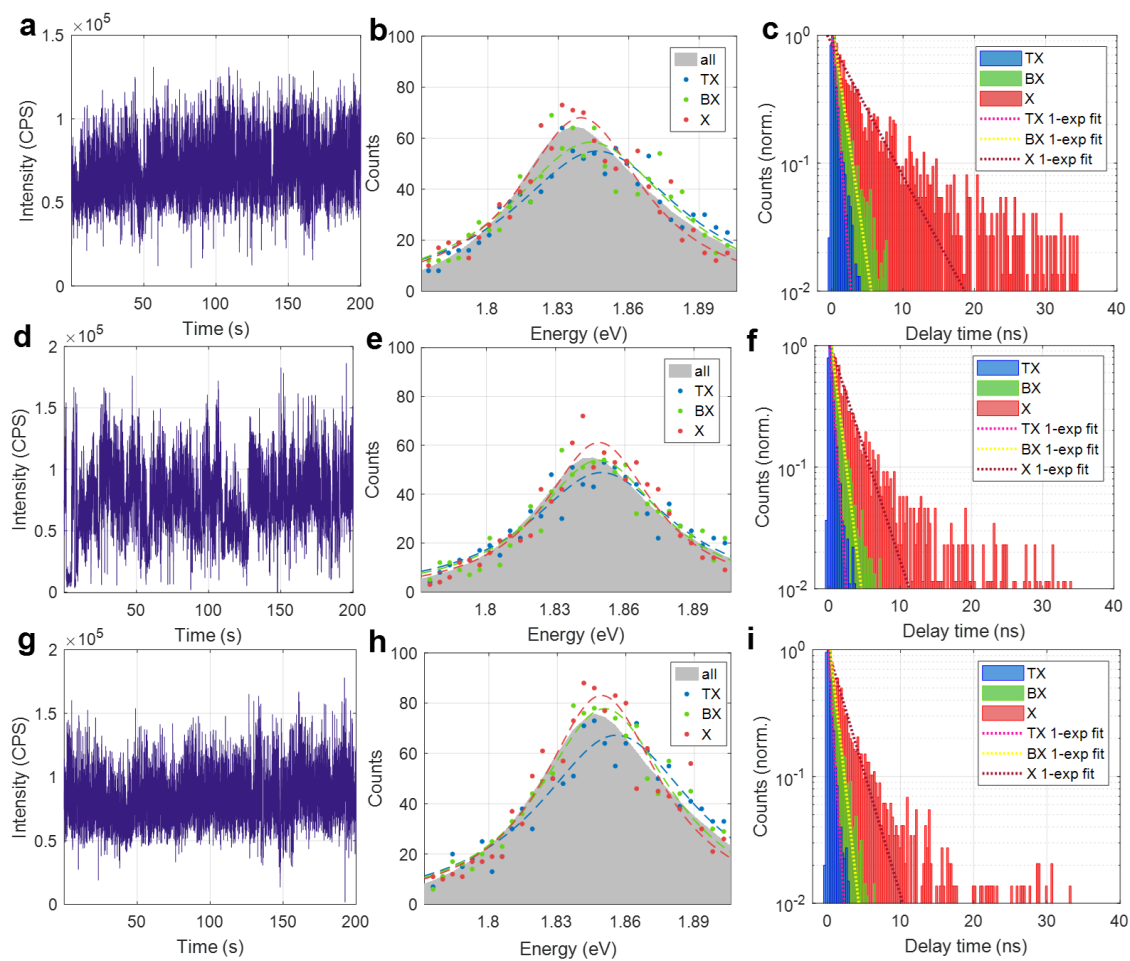

*Figure S6.* Other examples of single-NPL analysis. Panels a–c correspond to NPL1; Panels d–f correspond to NPL2; Panels g–i correspond to NPL3. (a,d,g) Intensity trace for a 200-seconds acquisition time window for the three different NPLs. (b,e,h) Single NPL spectra of the triexciton (blue), biexciton (green), and single exciton (red) photons for the three NPLs. The blue, green, and red dashed lines are fits of the triexciton, biexciton, and single exciton spectra, respectively, to a Cauchy-Lorentz distribution. The gray area corresponds to all detected photons, without heralded post-selection, normalized to match the triexciton counts. (c,f,i) Lifetime histograms of the three single NPLs along with monoexponential fits. The delay time of the triexciton is from the laser peak, and the delay times of the biexciton and single exciton are from the preceding arrival times of the triexciton and biexciton, respectively.

## S8: Second-order correlation versus time gating of photon detections[4]

The gated $g^{(2)}$ assay was done by gating the time delay of photons relative to the previous laser pulse and using only late-arriving photons to construct the $g^{(2)}(\tau)$ curve and calculate the gated $g^{(2)}(0)$. The chosen time gate applied for all NPLs is from 6 to 35 ns, meaning that only photons with time delays higher than 6 ns and lower than 35 ns are considered for the $g^{(2)}$ analysis. The lower limit of 6 ns aims to remove the contribution from multiexcitons, which relax faster, and the upper limit of 35 ns aims to remove noise. In this way, late-arriving photons should represent photons emitted from the single exciton state only, expected to increasing the degree of antibunching (reducing $g^{(2)}(0)$).[4,5] Figure S5 shows the degree of antibunching, the reduced probability of detecting two photons at the same laser period, versus gating times. Gating the photons at 3 ns already shows a decrease of the $g^{(2)}(0)$ below 0.5, which is taken as a criterion for a single emitter, meaning measuring only a single NPL at a time.[6]

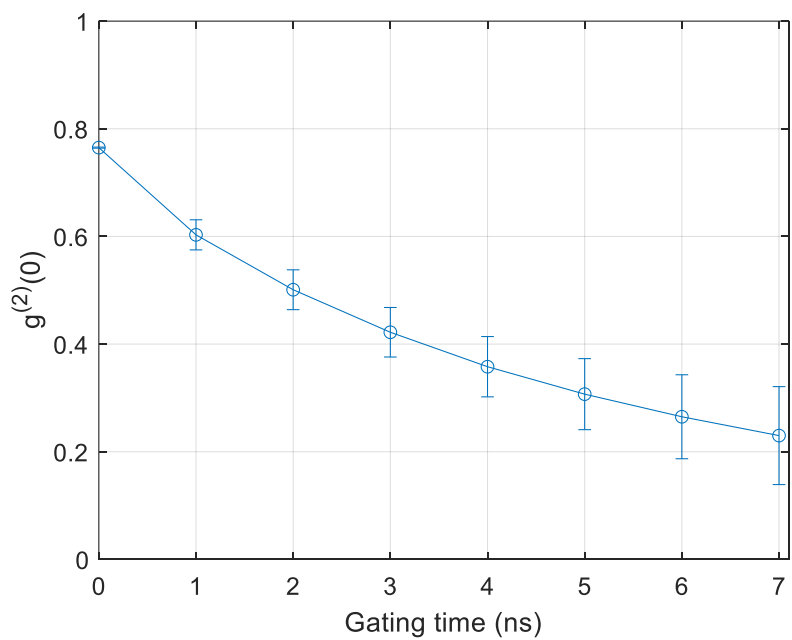

*Figure S7.* Second-order photon correlation values at zero time delay ($g^{(2)}(0)$) for increasing gating times in a single NPL. The $g^{(2)}(0)$ shows a decrease (i.e., increasing antibunching) with increasing the gating time, down to about 0.2 for a gating time of 7 ns.

# S9: Correlation of the triexciton and biexciton binding energies

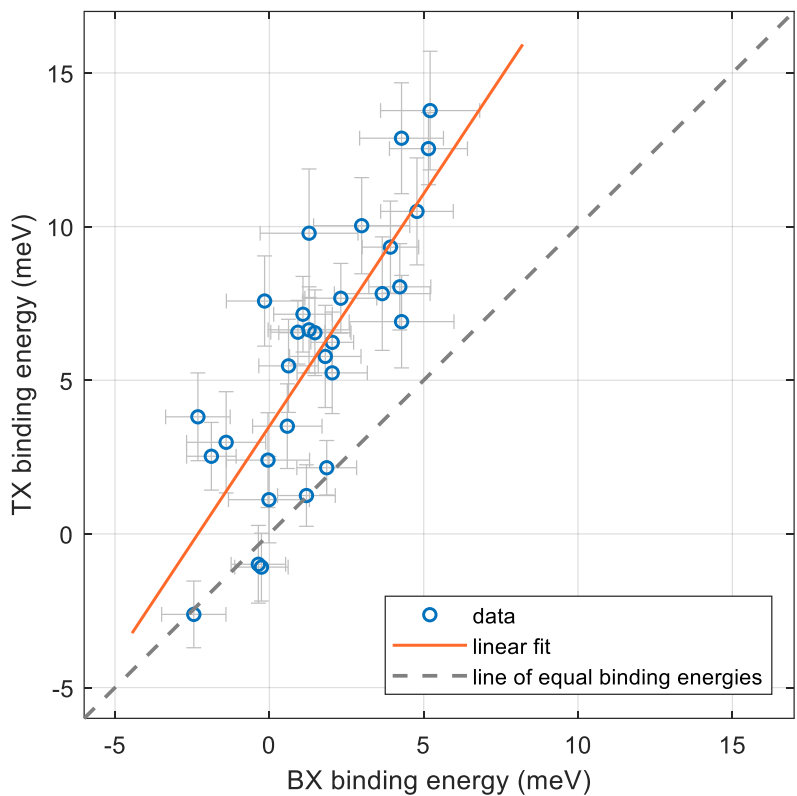

*Figure S8.* Correlation of the triexciton (TX) and biexciton (BX) binding energies of all measured single NPLs. The orange solid line is a linear fit to the data. The gray dashed line is a guide to the eye, indicating identical TX and BX binding energies.